\documentclass[a4paper, 11pt]{article}

\usepackage{a4wide}
\usepackage{amsmath}
\usepackage{latexsym}
\usepackage{graphicx}
\usepackage{color}
\usepackage[pdfstartview=FitH]{hyperref}
\usepackage{verbatim}
\usepackage{authblk}

\title{A multiplicative process for generating a beta-like survival function with application to the UK 2016 EU referendum results}

\author[1]{Trevor Fenner trevor@dcs.bbk.ac.uk}
\author[2]{Eric Kaufmann e.kaufmann@bbk.ac.uk}
\author[1]{\authorcr Mark Levene mark@dcs.bbk.ac.uk}
\author[1]{George Loizou george@dcs.bbk.ac.uk}
\affil[1]{Department of Computer Science and Information Systems}
\affil[2]{Department of Politics, Birkbeck, University of London \authorcr London WC1E 7HX, U.K.}

\date{}

\begin{document}

\maketitle

\begin{abstract}

Human dynamics and sociophysics suggest statistical models that may explain and provide us with better insight into social phenomena. Contextual and selection effects tend to produce extreme values in the tails of rank-ordered distributions of both census data and district-level election outcomes. Models that account for this nonlinearity generally outperform linear models. Fitting nonlinear functions based on rank-ordering census and election data therefore improves the fit of aggregate voting models. This may help improve ecological inference, as well as election forecasting in majoritarian systems.

We propose a generative multiplicative decrease model that gives rise to a rank-order distribution, and facilitates the analysis of the recent UK EU referendum results. We supply empirical evidence that the beta-like survival function, which can be generated directly from our model, is a close fit to the referendum results, and also may have predictive value when covariate data are available.

\end{abstract}


\section{Introduction}

Recent interest in complex social systems, such as social networks, the world-wide-web, email networks and mobile phone networks \cite{BARA07}, has led researchers to investigate the processes that could explain the dynamics of human behaviour within these networks. Human dynamics is not limited to the study of behaviour in communication networks, and has a broader remit similar to the aims of {\em sociophysics} \cite{GALA08,SEN14,STAU14}, which uses concepts and methods from statistical physics to investigate social phenomena, opinion formation and political behaviour. A central idea here is that, in the context of statistical physics, individual humans can be thought of as ``social atoms'', each exhibiting simple individual behaviour and possessing very limited intelligence, but nevertheless collectively yielding complex social patterns \cite{BENT11}. In sociophysics, critical phenomena are important in demonstrating how the transition to global behaviour can emerge from the interactions of many individual social atoms, for example, as described in \cite{PAN10}, where the popularity of movies emerges as collective choice behaviour.
The employed methodology often involves the empirical investigation of collective choice dynamics, resulting in the postulation of statistical laws governing, for example, universal properties of election results \cite{CHAT13}. The existence of such universal patterns, which are consistent across different countries and over a considerable time span, may suggest a fundamental underlying social pattern that could provide insight into the processes that determine human decision making \cite{FORT12}. A crucial issue in the process of statistical model building is that the model must be tested against experimental data and superseded by a newer model that better explains the data when such a model is found \cite{GALA08}.

\smallskip

In the context of human dynamics, we have been particularly interested in formulating {\em generative models} in the form of stochastic processes by which complex systems evolve and give rise to power laws or other distributions \cite{FENN05,FENN12,FENN15}. This type of research builds on the early work of Simon \cite{SIMO55}, and the more recent work of Barab\'asi's group \cite{ALBE01} and other researchers \cite{BORN07a}. In the bigger picture, one can view the goal of such research as being similar to that of {\em social mechanisms} \cite{HEDS98}, which looks into the processes, or mechanisms, that can explain observed social phenomena. Using an example given in \cite{SCHE98a}, the growth in the sales of a book can be explained by the well-known logistic growth model \cite{TSOU02}, and more recently we have shown that the process of conference registration with an early bird deadline can be modelled by bi-logistic growth \cite{FENN13}.

\smallskip

In this paper, we employ the multiplicative process \cite{MITZ04,ZANE08} introduced in \cite{FENN16b}, which is described  by the same underlying equations as the generative model proposed in \cite{FENN15}. This was designed to capture the essential dynamics of survival analysis applications \cite{KLEI12}.
As in \cite{FENN16b}, we introduce rank-ordering into the model as a natural mechanism in situations where there is no intrinsic ordering of the data, such as constituency-based election results \cite{FENN16b} or, as in this paper, regional referendum results. Rank-ordering \cite{SORN96} is a technique in which we rank the data objects according to some numerically described feature. We then plot this feature against rank, and finally analyse the resulting distribution. Examples of rank-order distributions are: the distribution of large earthquakes \cite{SORN96}, the distribution of oil reserve sizes \cite{LAHE98}, Zipf's rank-frequency distribution \cite[Section~1.4.3]{MANN99}, the size distribution of cities  \cite{BRAK99},
and the distribution of historical extreme events \cite{CHEN12}.

\smallskip

In many real-world situations, Zipf's distribution, or more generally a power-law distribution, may only manifest itself for small and intermediate ranks, while for larger ranks a more pronounced cutoff is observed \cite{MART09}.
This has led researchers to combine these two regimes into a single distribution, called {\em the beta-like function} \cite{NAUM08,MART09,AUSL16},  which appears to exhibit universal behaviour for rank-order distributions and is asymptotically a beta distribution \cite[Chapter 25]{JOHN95}. Whereas, in \cite{NAUM08}, the beta-like function is shown to be indirectly linked to a multinomial multiplicative process, here we introduce the {\em beta-like survival function} as a direct and intuitive consequence of a multiplicative decrease process, where an {\em attrition function} controls the rate of decrease of the survival function at each stage of the process. In Section~\ref{sec:beta} we will show that the beta-like function can be approximated by our beta-like survival function. In particular, the attrition function is a mixture of preferential (cf. preferential attachment \cite{ALBE01}) and uniform attrition mechanisms, where the former is inversely proportional to the rank of the voting district and the latter is uniform over the voting districts.

\smallskip

The main contribution of the paper is to demonstrate the suitability of the beta-like survival function $-$ a rank-ordered distribution generated from a multiplicative process $-$ for modelling the UK 2016 EU regional referendum results.
The EU referendum involved a binary choice, where voters had to choose between remaining or leaving the EU.
Voting behaviour in referenda often exhibits greater volatility than that found in general elections \cite{LEDU02}.
In a referendum, parties are often internally divided over the issues, so party ideology is less of an issue than in a general election. Moreover, uncertainties can introduce shifts in opinion when doubts are raised on important issues during the course of a campaign. The UK EU referendum, also known as the ``Brexit'' referendum, will be analysed for many years to come, as it is considered a momentous event in the history of Europe.


\smallskip

Since Robinson's seminal paper on the ecological fallacy \cite{ROBI50}, the use of aggregate data in political science has fallen into abeyance, superseded by survey methodology \cite{KING97}. Most election analyses therefore focus on individual-level data. However, the fact that different districts in a country vote differently is also important. First, election results represent a complete sample, unaffected by survey sampling bias. Second, electoral geography is an intrinsically interesting phenomenon, viz. the popular use of generalisations such as `red' and `blue' states or counties in America. Third, where individual-level data is unavailable, as is the case in historical analysis, for small parties, or for many elections, ecological inference is required \cite{KING97}. Finally, political scientists and practitioners pragmatically wish to know how the vote is dispersed across electoral districts as this can radically affect the number of seats a party wins in a majoritarian system. In the 2015 British general election, for instance, UKIP's dispersed 12.6\% vote gave it just one seat, while the Scottish Nationalist Party (SNP) won 56 seats with their concentrated 4.7\%. Likewise, in the 1994 elections for the House of Representatives in Iowa, the Democrats' dispersed 42\% of the vote failed to earn it a single one of the state's five House seats.

Aggregate models are also important for predicting election results in majoritarian systems. Compared to polls or surveys which include attitudinal predictors, socioeconomic specifications predict little of the variation in individual-level voting. Yet there is a quandary. At the aggregate level, census-based socioeconomic models account for much of the variation in seat-level outcomes. While scholars of voting downplay socio-demographics, these are enjoying a renaissance among election forecasters, who have augmented polls with census data to refine seat-level predictions \cite{ROMA08} \cite[Chapter 2]{SILV12}. These outperform blanket vote-to-seat conversion rules such as the {\em cube law} \cite{KEND50}. Panel studies deploying aggregate data, which consider the effects of variables such as economic change, ‘homegrown’ candidacy and incumbency on election results, perform extremely well in predicting seat-level changes over time \cite{KAHA09,FAIR09}. In this paper, we show how the use of a rank-ordered distribution can improve the accuracy of aggregate-level models. In particular, we focus on Britain's referendum on whether to leave the European Union. Held on 23 June 2016, the Leave side won an unexpected 52-48 percent victory.

\medskip

The rest of the paper is organised as follows.
In Section~\ref{sec:urn} we present a multiplicative process that provides us with a mechanism to model the essential dynamics of ranked-ordered models, and we derive a well-known differential equation to describe the process.
In Section~\ref{sec:eu-ref} we show how referendum results can be modelled using the multiplicative process presented in Section~\ref{sec:urn}, via the rank-ordering method.
In Section~\ref{sec:beta} we introduce the beta-like survival function, which is then used in Section~\ref{sec:uk} to model the results of the UK 2016 EU referendum.
Finally, in Section~\ref{sec:conc} we give our concluding remarks.

\section{A multiplicative process for generating a rank-order distribution}
\label{sec:urn}

We next present a generative model in the form of a {\em multiplicative process} \cite{MITZ04,ZANE08} that can also be viewed as a survival model similar to the one introduced in \cite{FENN15}, in the context of human dynamics; this model was first introduced in \cite{FENN16b} but we repeat it in the context of a referendum for completeness.
In its simplest form, a multiplicative process generates a log-normal distribution \cite[Chapter 14]{JOHN94}, \cite{LIMP01}, and has applications in many fields, such as economics, biology and ecology \cite{MITZ04}.
The solution to the multiplicative process we propose will be utilised in Section~\ref{sec:eu-ref}, in the context of a rank-ordered model of the proportion of votes attained for a particular answer in a multiple-choice question referendum.

\smallskip

We assume a countable number of indices where, for a given answer, the $i$th index represents the $i$th district ranked in descending order of the number of votes for that answer. For any stage $s$, $s \ge 0$, we let $\mu(i,s)$, $0 \le \mu(i,s) \le 1$, be the probability that a potential vote is lost in the $i$th district at that stage. Usually $\mu(i,s)$ is known as the {\em mortality rate function}, but here we prefer to call it the {\em attrition function}, which is more descriptive in the context of voting. We always require that $\mu(0,s)=0$ for all $s$.

\smallskip

We now let $F(i,s)$, $0 \le F(i,s) \le 1$, be a discrete function representing, for a given answer, the expected proportion of the popular vote potentially attainable for that answer in district $i$ at stage $s$. Initially, we set $F(0,0) = 1$ for a dummy district $0$, and $F(i,0) = 0$ for all $i > 0$.

\smallskip

The dynamics of the multiplicative process can be captured by the following two equations:
\begin{equation}\label{eq:init}
F(0,s) = 1 \ \ {\rm for} \ \ s \ge 0,
\end{equation}
and
\begin{equation}\label{eq:diff}
F(i+1,s+1) = \left(1 - \mu(i,s)\right) F(i,s) \ \ {\rm for} \ \ 0 \le i \le s.
\end{equation}
\smallskip

Equations (\ref{eq:init}) and (\ref{eq:diff}) define the expected behaviour of a stochastic process \cite{ROSS96} describing how, as $i$ increases, the vote decreases in districts where the given answer is less popular. For any particular vote, the attrition function is the probabilistic mechanism that decides whether the vote will be lost or not. The process obeys Gibrat's law \cite{EECK04}, which in its original form states that the proportional rate of growth of a firm is independent of its absolute size. In our context, Gibrat's law states that the proportional rate of decrease in the popular vote is independent of the actual number of votes cast for the answer in the district.

\medskip

As in \cite{FENN15}, we approximate the discrete function $F(i,s)$ by a continuous function $f(i,s)$, and $\mu(i,s)$ is now also a continuous function; $f(i,s)$ is known as the {\em survival function}. Initially, we have $f(0,s)= 1$ for all $s$, and $f(i,0) = 0$ for all $i > 0$.

\smallskip

The dynamics of the model is now captured by the first-order hyperbolic partial differential equation \cite{LAX06},
\begin{equation}\label{eq:partial}
\frac{\partial f(i,s)}{\partial s} + \frac{\partial f(i,s)}{\partial i} + \mu(i,s) f(i,s) = 0,
\end{equation}
which is the same as that encountered in age-structured models of population dynamics \cite{CHAR94}.

\smallskip

Equation (\ref{eq:partial}) is the well-known {\em transport equation} in fluid dynamics \cite{LAX06}, and
the {\em renewal equation} in population dynamics \cite{PILA91,CHAR94,LI08}.
Following Equation $1.22$ in \cite{CHAR94}, the solution of (\ref{eq:partial}), when $i \le s$, is given by
\begin{equation}\label{eq:renewal}
f(i,s) = \exp \left( - \int_{0}^{i} \mu \left( i-t, s-t \right) dt \right).
\end{equation}

\section{Application of the model to analysis of referendum results}
\label{sec:eu-ref}

We consider a multiple-choice question referendum with several options, where the votes are counted over the whole electorate \cite{LEDU02}. Often, as in the case of the UK EU referendum, only two choices are presented to the public: accepting or rejecting a proposal. In a referendum, unlike in an election, voters are not choosing among political parties or political candidates, and thus campaigns are often cross-party, as in the UK EU referendum. The voting takes place across the country in a designated number of Local Authority districts, and the votes are aggregated country-wide.

\smallskip

We make use of the rank-ordering technique \cite{SORN96} in the context of a {\em voter model} \cite{FENN16b} as follows.
Focusing on one particular answer to the referendum question, we model the proportion of votes $V_i$ attained for that answer in district $i$, where  $i$ represents the {\em rank} of the district and $0 \le V_i \le 1$. Thus, ordering the districts in descending order of the proportion of votes, we obtain the {\em votes vector} $(V_i) = (V_0, V_1, \cdots)^{T}$, where:
\begin{equation}\label{eq:votes}
V_0 > V_1 > V_2 > \cdots V_i > \cdots.
\end{equation}
\smallskip

Dsitrict $0$ is a ``dummy'' district with $V_0 = 1$. In the unlikely event that two districts have exactly the same proportion of votes, their order is chosen randomly.

\medskip

The votes vector is analogous to the empirical survival function $\hat{S}(\cdot)$ \cite{KLEI12}, where $V_i$, which corresponds to $\hat{S}(i)$, can be viewed as an estimate of the expected proportion of the popular vote in district $i$, given that $V_{i-1}$ was the proportion in district $i-1$; cf. the {\em Kaplan-Meier estimator} \cite{KAPL58,KLEI12} in the context of survival models. In the context of the voter model, we see that $\hat{S}(i) \approx f(i,s^*)$, where $s^*$, the final stage, is equal to the number of voting districts.

\smallskip

The rank-ordering of the districts, as in (\ref{eq:votes}), can be simulated by the multiplicative process described in Section~\ref{sec:urn}, where $i$ corresponds to the $i$th highest ranked district. An appropriate attrition function $\mu(i,s)$ is used, which is often decreasing in $i$. In terms of the voter model, as we consider less popular districts for the given answer, i.e. those of lower ``rank'' (remembering that a lower rank is represented by a higher seat number), more votes are lost.

\smallskip

Thus, given a suitable attrition function, as presented in the next section, we can generate the votes vector using the process described in Section~\ref{sec:urn}. In the next section, we propose the beta-like survival function as a possible distribution for the votes vector, and in Section~\ref{sec:uk} we demonstrate its suitability for modeling the 2016 UK EU referendum.

\section{The beta-like survival function}\label{sec:beta}

The {\em beta-like function} \cite{NAUM08,MART09}, see (\ref{eq:beta-like}) below, is a discrete version of the beta distribution \cite[Chapter 25]{JOHN95}, which has been shown to be a very good fit for a variety of rank-ordered data distributions. Here we propose the similar {\em beta-like survival function}, which is derived from (\ref{eq:renewal}) with the specific attrition function introduced in (\ref{eq:mixture}) below. In \cite{NAUM08}, an argument is given that relates the beta-like function to a multinomial multiplicative process, while here we show that it can be derived as a direct consequence of the multiplicative process introduced in Section~\ref{sec:urn} with the attrition function in (\ref{eq:mixture}).

\smallskip

We now derive the beta-like survival function from (\ref{eq:renewal}) by introducing the following attrition function, which is the mixture of rank-dependent and rank-independent attrition,
\begin{equation}\label{eq:mixture}
\mu(i,s) = \frac{\alpha}{i+\kappa} + \frac{\beta}{s},
\end{equation}
where $\alpha, \beta$ and $\kappa$ are positive constants, and $i \le s$.

\medskip

In the context of the generative model introduced in Section~\ref{sec:urn},
the rank-dependent component $\alpha/(i+\kappa)$ in (\ref{eq:mixture}) models {\em preferential attrition}, i.e. the potential loss of a vote is dependent on the rank $i$ of the district, where $i \le s$. On the other hand, the rank-independent component $\beta/s$ models {\em uniform attrition} at stage $s$, where we note that $s$ is bounded below by the number of voting districts.

Preferential attrition might occur, for instance, because a district's rank (i.e Brexit vote share) may exert a contextual effect on the voting decisions of its constituent individuals; or may attract movers from other districts with similar political characteristics; or may produce an increased supply of local election volunteers. All produce positive feedbacks.

\smallskip

Thus, from (\ref{eq:renewal}), it follows that the survival function is given by
\begin{align}\label{eq:beta-like-weibull}
f(i,s) &= \exp \left( \alpha \ln \left(\frac{\kappa}{i + \kappa} \right) + \beta \ln \left( 1 - \frac{i}{s}\right) \right) \nonumber \\
       &= \left( \frac{\kappa}{i + \kappa} \right)^\alpha \left( 1 - \frac{i}{s} \right)^\beta.
\end{align}

We call $f(i,s)$ the {\em beta-like survival function}, motivated by the following argument.

\medskip

Letting $K=\kappa^\alpha s^{-\beta}$ and assuming that $\kappa$ is much smaller than $i$, equation~(\ref{eq:beta-like-weibull}) can be approximated by
\begin{equation}\label{eq:beta-like}
f(i,s) \approx K i^{-\alpha} \left( s - i \right)^\beta,
\end{equation}
which is the beta-like function proposed in \cite{NAUM08}; we note that we obtain the {\em Lavalette function} in the special case when $\alpha = \beta$ \cite{FONT16}.

Therefore, the multiplicative process presented in Section~\ref{sec:urn} provides a direct and intuitive generative model for the beta-like function. In contrast, the multinomial multiplicative process described in \cite{NAUM08} is indirectly linked to the beta-like function via a stretched exponential \cite{LAHE98} derived by ranking the components of a multinomial distribution, and only afterwards fitting a beta-like function.

\medskip

We crystalise (\ref{eq:beta-like-weibull}) by fixing $\kappa=0.5$ and including a scaling constant $C$ for normalisation purposes, so that
\begin{equation}\label{eq:beta-like-kappa}
f(i,s) = C \left( \frac{0.5}{i + 0.5} \right)^\alpha \left( 1 - \frac{i}{s} \right)^\beta.
\end{equation}

The justification for fixing $\kappa$ is that its sole purpose in (\ref{eq:mixture}) is to prevent the first term being undefined when $i = 0$; setting $\kappa = 0.5$ seems sensible since, when $i$ is large, the precise value of $\kappa$ is rather unimportant.


\smallskip

It can be seen that the beta-like survival function in (\ref{eq:beta-like-kappa}) combines two regimes. When $i$ is close to $s$, the second term dominates exhibiting a polynomial decay, otherwise the first term dominates exhibiting power-law behaviour. In terms of (\ref{eq:mixture}), the preferential attrition component gives rise to the power-law regime, while the uniform attrition component gives rise to the polynomial decay regime.

\smallskip

We define a linear transformation $f^*(i,s)$ of $f(i,s)$ as follows:
\begin{equation}\label{eq:beta-like-linear}
f^*(i,s) = \tau f(i,s) + \rho = \tau  \left( \frac{0.5}{i + 0.5} \right)^\alpha \left( 1 - \frac{i}{s} \right)^\beta + \rho,
\end{equation}
where the scaling constant $C$ from (\ref{eq:beta-like-kappa}) is absorbed into the {\em slope} parameter $\tau$, and $\rho$ is the {\em shift} parameter.

\section{Analysis of the UK 2016 EU referendum results}
\label{sec:uk}

We now make use of the beta-like survival function, derived in Section~\ref{sec:beta}, to analyse the Remain and Leave votes of the 2016 EU referendum for the 382 Local Authority districts in the UK (we will often refer to these simply as districts); the full set of electoral results is available online at \cite{ELEC16}. All the computations were carried out using the Matlab software package.

\smallskip

As noted earlier, Britain voted to leave the European Union on 23 June 2016. This result surprised pollsters and commentators. Aggregate analysis subsequently showed that education was the strongest socio-demographic predictor of the vote \cite{GOOD16a}. Multi-level analysis of individual data revealed that this was mainly due to the compositional effect of less-educated people voting Leave, but also because of a contextual effect, whereby those with a degree living in areas with lower average education tended to have their opinions shaped by their community and vice-versa. As Goodwin and Heath write, those with low qualifications were 16 points less likely to vote to leave the EU if they lived in an area with the highest as opposed to lowest average education in the country. Those with A-levels or a degree living in these high-skill areas were 30 points less likely to have voted to leave than their similarly qualified compatriots living in the lowest-educated areas \cite{GOOD16b}. The latter process introduces a positive feedback between aggregate and individual voting: aggregates shape individuals, who in turn comprise aggregates. These positive feedbacks help explain electoral extremes, such as the 78.6\% of the London borough of Lambeth who voted Remain or the 75.6\% in Boston, Lincolnshire who opted to Leave. This nonlinear clustering distorts linear regression results, producing the inflated model fits noted in Robinson's \cite{ROBI50} article. In many elections, local campaign effects are likewise important: areas with particular social characteristics tend to be targeted or avoided by parties, and tend to produce a larger or smaller supply of local volunteers. Populist right parties, for example, focus their resources on whiter, less educated districts, which also produce more volunteers. We know from the British Election Study that, when people are contacted by a party, they are more likely - all other things being equal - to have voted for that party \cite{FIEL16}. Thus contextual effects enhance political supply, which contributes to further positive feedbacks and nonlinear vote distributions over location. Taking account of this clustering within the geographic distribution of both independent and dependent variables helps improve model fit in aggregate election analyses.

\smallskip

In Subsection~\ref{subsec:referendum}, using nonlinear least squares regression, we fit beta-like survival functions $f(i,s)$ to the referendum results vectors $(V_i)$ for both the entire UK and Scotland, and in Subsection~\ref{subsec:covariates} we investigate how the beta-like survival function can be used to associate regional covariates with the referendum data.
We use subscripts $U$ and $S$ to denote the UK and Scotland, respectively, for example, $f_U(i,s)$ and $f_S(i,s)$.

\subsection{Analysis of the referendum results using the beta-like survival function}
\label{subsec:referendum}

The overall result for the entire UK electorate was 48.11\% for Remain and 51.89\% for Leave. The fitted parameters $\alpha$, $\beta$ and $C$ for the UK regional referendum results as a whole, together with the coefficient of determination $R^2$ \cite{MOTU95} are shown in Table~\ref{table:ref1}. As can be seen, the power-law exponent $\alpha$ for Leave is significantly lower than that for Remain, while the decay exponent $\beta$ for Leave is somewhat higher than that for Remain. This may indicate that the proportions of votes for Leave were more ``stable'' across the country than those for Remain.
In other words, it is feasible that positive feedbacks driven by contextual effects on individual vote choice mattered more in Remain than Leave areas. Both $R^2$ values are very high, indicating very good fits of the beta-like survival function to the data.

\begin{table}[ht]
\begin{center}
\begin{tabular}{|l|c|c|c|c|}\hline
Option  & $\alpha$ & $\beta$ & $C$    & $R^2$  \\ \hline \hline
Leave   & 0.0357   & 0.2094  & 0.7801 & 0.9913 \\ \hline
Remain  & 0.1286   & 0.1244  & 1.0740 & 0.9930 \\ \hline
\end{tabular}
\end{center}
\caption{\label{table:ref1} Nonlinear least-squares regression fitting the beta-like survival function $f_U(i,s)$ to the empirical results vector $(V_i)_U$ of the regional results for the UK EU referendum.}
\end{table}

\begin{figure}[ht]
\begin{minipage}{0.5\textwidth}
\includegraphics[scale=0.39]{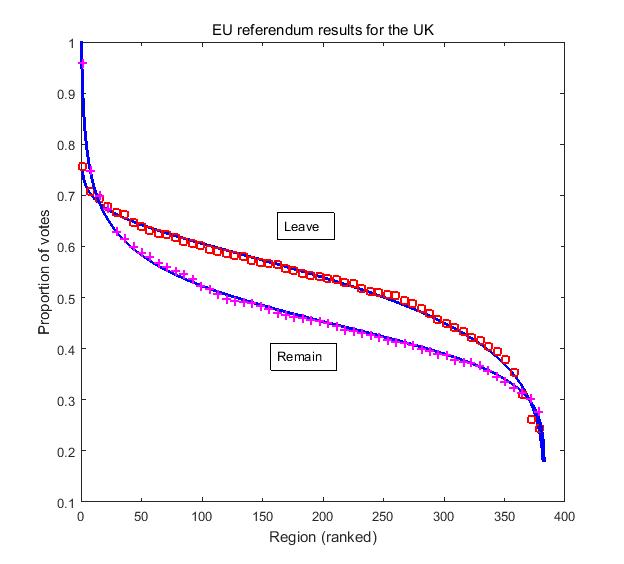}
\end{minipage}
\begin{minipage}{0.5\textwidth}
\includegraphics[scale=0.39]{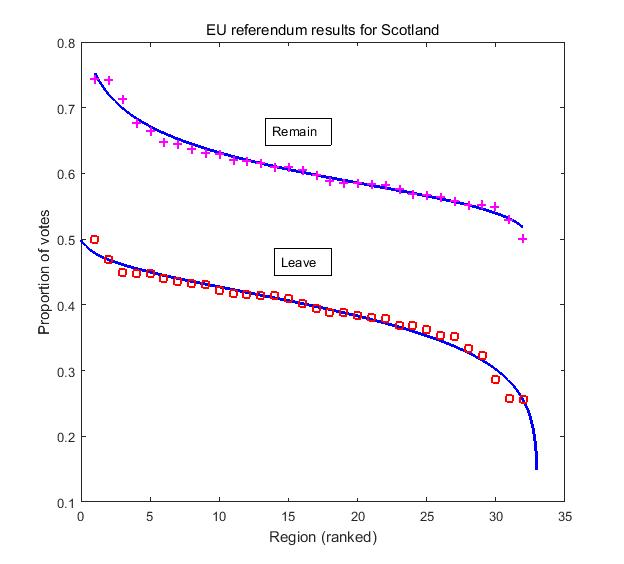}
\end{minipage}
\caption{\label{figure:results} Regression curve and regional data points for the UK (left) and for Scotland (right).}
\end{figure}

We now consider the results over the 32 Local Authority districts in Scotland, which, in contrast to the UK as a whole, had a majority of 62\% for Remain. The fitted parameters for Scotland are shown in Table~\ref{table:scot-ref1} together with the $R^2$ values,
which again indicate very good fits. As for the overall UK results, the proportions of votes for Leave were more ``stable'' than those for Remain, despite the different overall result.
The fitted curves and data points for the districts, for both the entire UK and Scotland, are shown graphically in Figure~\ref{figure:results}.

\begin{table}[ht]
\begin{center}
\begin{tabular}{|l|c|c|c|c|}\hline
Option  & $\alpha$ & $\beta$ & $C$    & $R^2$  \\ \hline \hline
Leave   & 0.0322   & 0.1540  & 0.4984 & 0.9765 \\ \hline
Remain  & 0.0848   & 0.0333  & 0.8274 & 0.9822  \\ \hline
\end{tabular}
\end{center}
\caption{\label{table:scot-ref1} Nonlinear least-squares regression fitting the beta-like survival function $f_S(i,s)$ to the empirical vector $(V_i)_S$ of the regional results for the Scottish EU referendum.}
\end{table}
\medskip

\smallskip

We further compare the regional patterns for the UK and Scotland by introducing linear transformations $f^*(i,s)$ of the beta-like survival functions, as in (\ref{eq:beta-like-linear}), using the  fitted parameters $\alpha$, $\beta$ and $C$ given in Tables \ref{table:ref1} and \ref{table:scot-ref1}. Using least squares approximation, we then fit the transformed function $f^*_S(i,s)$ for Scotland to the referendum results vector $(V_i)_U$ for the UK. The resulting {\em shift} and {\em slope} parameters, $\rho$ and $\tau$, together with the $R^2$ values are shown in Table~\ref{table:scot-uk}. Similarly, Table~\ref{table:uk-scot} shows the result of fitting $f^*_U(i,s)$ to $(V_i)_S$. The linear transformations demonstrate that although the overall results for the UK as a whole and Scotland were very different, as can be seen in Figure~\ref{figure:results}, the patterns for the regional proportions for the UK and Scotland are linearly related for both Leave and Remain.
Note that in both functions there are extreme results which suggest that positive feedbacks are operating, whether through contextual effects (i.e. \cite{GOOD16b}) or selective migration of those with pro-Leave or pro-Remain characteristics toward districts where they are already concentrated.

\begin{table}[ht]
\begin{center}
\begin{tabular}{|l|c|c|c|}\hline
Data   & {\em shift} & {\em slope} & $R^2$  \\ \hline \hline
Leave  & -0.1538     & 0.9428      & 0.9937 \\ \hline
Remain & -0.4935     & 1.5993      & 0.9871 \\ \hline
\end{tabular}
\end{center}
\caption{\label{table:scot-uk} The {\em shift} and {\em slope} parameters, $\rho$ and $\tau$, fitting $f^*_S(i,s)$ to $(V_i)_U$.}
\end{table}

\begin{table}[ht]
\begin{center}
\begin{tabular}{|l|c|c|c|}\hline
Data   & {\em shift} & {\em slope} & $R^2$  \\ \hline \hline
Leave  & 0.0876      & 0.4092      & 0.9941 \\ \hline
Remain & 0.3213      & 0.4853      & 0.9742 \\ \hline
\end{tabular}
\end{center}
\caption{\label{table:uk-scot} The {\em shift} and {\em slope} parameters, $\rho$ and $\tau$, fitting $f^*_U(i,s)$ to $(V_i)_S$.}
\end{table}

\subsection{Analysis of four covariates with the beta-like survival function}
\label{subsec:covariates}

We now outline the methodology of how covariates may be used in association with the referendum results.
The dependent variable for this stage is Leave vote share in a Local Authority district. This is regressed on 2011 census data \cite{ONS13}, also denominated in percentages.


The most common method employed for such an analysis is linear regression, although generalised linear models are recommended when the distribution of the dependent variable may not be normal and may not vary linearly with the independent variables \cite{FREU06}.

\smallskip

To demonstrate how our rank-order distribution may be used for explanatory purposes, we outline a baseline methodology using a single covariate. Our methodology is as follows, where $\gamma$ denotes a covariate.

\renewcommand{\labelenumi}{(\roman{enumi})}
\begin{enumerate}
\item We first order $\gamma_i$, for districts $i=0,1,2,\ldots$, in descending order, where $\gamma_0=1$ for the dummy district $0$, in order to obtain the empirical coavriate vector $(\gamma_i)$.
\item We then use nonlinear least-squares regression to fit a beta-like survival function $g(i,s)$ to the vector $(\gamma_i)$ from (i); this gives the fitted parameters $\alpha$, $\beta$ and $C$, as in (\ref{eq:beta-like-kappa}).
\item We now use linear regression to fit the transformed beta-like survival function $g^*(i,s)$, with the values of $\alpha$, $\beta$ and $C$ obtained in (ii), to the votes vector $(V_i)$ of the Leave results of the referendum. This yields the corresponding {\em shift} and {\em slope} parameters, $\rho$ and $\tau$, respectively, as in (\ref{eq:beta-like-linear}). Since voters had only two choices, we could equivalently choose to regress on the Remain results.
\end{enumerate}
\smallskip

As a proof of concept we chose four census covariates, which the literature suggests may be associated with the Leave vote:
`White-qualification', which represents the average qualification level of the white British population in a district (excluding Scotland); `Identify-as-English', which represents the share of White British people in a district identifying as English rather then British, Irish or Welsh (excluding Scotland and Wales); `Social-grade', which represents the average occupational level of the White British population in a district (excluding Scotland); and `Carstairs-index', which represents the Carstairs deprivation index in a district (excluding Scotland) \cite{MORG06}. The Carstairs index of multiple deprivation, developed by Paul Norman, is an index of four components from the census. Namely, share of: residents without cars, male unemployed, low status occupational groups and overcrowded households.
Social grades (AB, C1, C2, DE) and qualification levels (none, 1, 2, apprenticeship, 3, 4 and above) were coded and averaged into an index for each Local Authority district.

\smallskip

\begin{figure}[ht]
\centering{\includegraphics[scale=0.6]{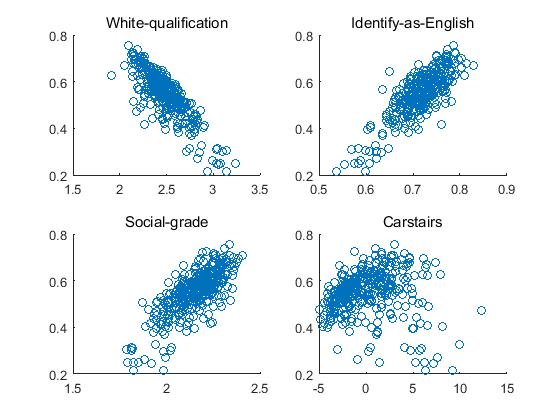}}
\caption{\label{figure:scatter} Scatter plots for the four chosen covariates; the y-values represent the proportion of Leave votes for a district and the x-values represent the values of the covariate for the district.}
\end{figure}

As an exploratory step, we show, in Figure~\ref{figure:scatter}, a scatter plot of the proportion of votes against the covariate values. These exhibit strong correlations for all the covariates apart from the Carstairs-index. The actual Pearson and Spearman correlations and the linear regression parameters, i.e. the {\em shift} and {\em slope}, are given in Table~\ref{table:corr}. We note that White-qualification has a negative correlation with the proportion of Leave votes, and that for the
Carstairs-index a weighted least squares linear regression \cite{FREU06} was applied to obtain the {\em shift} and {\em slope}. We further note that, as expected, the $R^2$ value for the Carstairs-index is significantly lower than for the others, even though the $R^2$ values for the other three covariates are not particularly high.

\begin{table}[ht]
\begin{center}
\begin{tabular}{|c|c|c|c|c|c|}\hline
Covariate           & Pearson & Spearman & {\em shift} & {\em slope} & $R^2$  \\ \hline \hline
White-qualification & -0.8340 & -0.7876  & 1.5335      & -0.3971     & 0.6955 \\ \hline
Identify-as-English &  0.8139 &  0.7375  & -0.6730     &  1.7063     & 0.6624 \\ \hline
Social-grade        &  0.7602 &  0.7171  & -0.7904     &  0.6226     & 0.5779 \\ \hline
Carstairs-index     &  0.0267 &  0.2664  &  0.5639     &  0.0120     & 0.4161 \\ \hline
\end{tabular}
\end{center}
\caption{\label{table:corr} Pearson and Spearman correlation between the proportion of votes and the covariate values per district, and the parameters obtained from their linear regression.}
\end{table}

\begin{table}[ht]
\begin{center}
\begin{tabular}{|l|c|c|c|c|c|}\hline
Covariate            & $\alpha$ & $\beta$ & $C$    & $R^2$  \\ \hline \hline
White-qualification  & 0.0548   & 0.0370  & 3.4830 & 0.9964 \\ \hline
Identify-as-English  & 0.0200   & 0.0559  & 0.8414 & 0.9966 \\ \hline
Social-grade         & 0.0189   & 0.0441  & 2.4870 & 0.9950 \\ \hline
Carstairs-index      & 0.1620   & 0.6626  & 19.860 & 0.9972 \\ \hline
\end{tabular}
\end{center}
\caption{\label{table:covariates} Nonlinear least-squares regression fitting of a beta-like survival function to the empirical  survival function of the four covariates.}
\end{table}

In Table~\ref{table:covariates} we see the fitted parameters to the beta-like survival function for each of the four covariates together with their $R^2$ values, which indicate a very good fit for all covariates.

In Table~\ref{table:linear} we give the fitted parameters for the linear transformation according to (\ref{eq:beta-like-linear}), together with their $R^2$ values, which indicate a very good fit for all covariates, apart from the Carstairs index where $R^2$ is less than $0.9$.
We observe from Table~\ref{table:linear} that the $R^2$ values are much higher than the ones in Table~\ref{table:corr}, indicating that our methodology using beta-like survival functions may yield better predictive models than traditional ones based on linear regression of the raw covariate data.

\begin{table}[ht]
\begin{center}
\begin{tabular}{|l|c|c|c|}\hline
Covariate           & {\em shift} & {\em slope} & $R^2$  \\ \hline \hline
White-qualification & -0.7267     & 1.6540      & 0.9911 \\ \hline
Identify-as-English & -0.9493     & 1.7610      & 0.9920  \\ \hline
Social-grade        & -1.2060     & 2.0300      & 0.9906  \\ \hline
Carstairs-index     &  0.3913     & 0.5879      & 0.8576  \\ \hline
\end{tabular}
\end{center}
\caption{\label{table:linear} The {\em shift} and {\em slope} coefficients from the linear transformation of the beta-like functions of the four convariates to the empirical survival function of the referendum results.}
\end{table}
\smallskip


Future research could examine how the deployment of beta-like survival functions of the form we have outlined might be used to generate nonlinear predictive models that could yield superior election predictions to the linear regression models currently used by election forecasters \cite{ROMA08} \cite[Chapter 2]{SILV12}.

\section{Concluding remarks}
\label{sec:conc}

Most phenomena in the social sciences are not normally distributed across geographical units because individuals and contexts influence each other. Contextual and selection effects lead to positive feedback loops which produce extreme geographic concentrations of both social characteristics and political opinions/behaviour.

We have proposed a multiplicative process that generates the rank-order distribution of the UK EU regional referendum results. In our model, the proportion of votes attained for a given party decreases according to a specified function, leading to a rank-order distribution represented by the votes vector. A solution to the continuous approximation of the equation specifying the model was given in (\ref{eq:renewal}), and is consequently identical to that of the renewal equation in population dynamics \cite{CHAR94}.

\smallskip

We suggest that nonlinear social models of aggregate voting behaviour can outperform linear models.
The beta-like survival function, obtained using an attrition function that is a mixture of preferential and uniform attrition mechanisms - thus generates an improved model for the UK EU referendum results.
Our results fit in well with the results in \cite{MART09}, where the beta-like function was shown to exhibit universal behaviour for rank-order distributions in several apparently unrelated disciplines.

\smallskip

We have also shown that the beta-like survival function could be instrumental in building a predictive model of the referendum results through a judicious choice of covariates. The methodology we presented in Subsection~\ref{subsec:covariates} may be used in tandem with traditional regression methods \cite{FREU06}, and may in fact have some advantages as rank-order distributions are very good at smoothing the data.

\smallskip

More research needs to be done on the methodology for working with rank-order dis¬tributions, and in particular with the beta-like survival function, due to its hypothesised universality. In particular, devising principled regression methods that combine the covari¬ates through multiple regression \cite{FREU06} would be useful. It is also worth considering other transformations, apart from the linear transformation in (\ref{eq:beta-like-linear}), which could potentially improve the predictive power of the model.
Aggregate models of voting are important for understanding electoral geography, inferring individual-level relationships in the absence of individual data and for predicting election results in majoritarian systems. Our nonlinear modelling technique, based on rank-ordering outcome and predictor variables, helps advance scholarship in these areas of political science.



\newcommand{\etalchar}[1]{$^{#1}$}

\end{document}